\documentstyle[eqsecnum,aps]{revtex}

\newcommand{\f}{\varphi}
\newcommand{\G}{G_{\mu \nu}}
\newcommand{\g}{g_{\mu \nu}}
\newcommand{\gu}{g^{\mu \nu}}

\begin{document}

\draft

\title{Nucleosynthesis Constraints on Scalar-Tensor Theories of Gravity}

\author{ D. I. Santiago\thanks{david@spacetime.stanford.edu}}
\address{ Department of Physics, Stanford University }
\author{ D. Kalligas} 
\address{ Physics Department, University of Athens }
\author{ R. V. Wagoner\thanks{wagoner@leland.stanford.edu}}
\address{ Department of Physics, Stanford University}
\date{\today}

\maketitle

\begin{abstract}
We study the cosmological evolution of massless single-field scalar-tensor 
theories of gravitation from the time before the onset of $e^+e^-$ annihilation
and nucleosynthesis up to the present. The cosmological evolution together with
the observational bounds on the abundances of the lightest elements (those
mostly produced in the early universe) place constraints on the coefficients
of the Taylor series expansion of $a(\f)$, which specifies the coupling of the
scalar field to matter and is the only free function in the theory. In the
case when $a(\f)$ has a minimum (i.e., when the theory evolves towards general
relativity) these constraints translate into a stronger limit on the
Post-Newtonian parameters $\gamma$ and $\beta$ than any other observational
test. Moreover, our bounds imply that, even at the epoch of annihilation and
nucleosynthesis, the evolution of the universe must be very close to that
predicted by general relativity if  we do not want to over- or underproduce
$\ ^{4}$He. Thus the amount of scalar field contribution to gravity is very
small even at such an early epoch. 
\end{abstract}


\section{Introduction}
\label{sec-int}

Scalar-tensor (ST) theories of gravitation have a long history. The first ones
to appear were those of Jordan\cite{jr}, Fierz\cite{fr}, and Brans-Dicke\cite
{bd}. These are the simplest theories in the sense that they consist of just 
one massless scalar field and its coupling strength to matter is constant. 
Later  Bergmann\cite{brg}, Nordtvedt\cite{ndt}, and Wagoner\cite{wag1} 
generalized the theory. In their versions the scalar field has a dynamic 
(scalar field dependant) coupling to matter and/or an arbitrary 
self-interaction. More recently, in a very comprehensive work,  Damour and 
Esposito-Far\`{e}se\cite{dm-ef} considered the theory with multiple scalar 
fields.

One could say that scalar-tensor theories of gravity are the simplest
alternative to Einstein's general theroy of  relativity (GR). Most recent
attempts at unified models of fundamental interactions, i.e.,string theories
\cite{str1}, predict the existence of scalar partners to the tensor gravity
of GR, and would have ST gravity as their low energy limit. In addition, these 
theories are important in inflationary cosmology where they provide a
way of exiting the inflationary epoch in a non-fine-tuned way\cite{infl}.
Scalar-tensor theories provide an important framework for comparison with
results of gravitational experiments that are going to take place in the
near future, e.g Gravity Probe-B\cite{gpb}, LIGOS\cite{lgo}, Geo\cite{geo},
and VIRGO\cite{vrg}.

In the present paper we study the cosmological evolution of the metric and
scalar gravitational fields in ST gravity. We consider the theories where  the
scalar field is massless, i.e., no self interaction. Although it seems natural
that without a gauge invariance the scalar would acquire a mass, certain 
results in string theory suggest that massless scalars arise naturally\cite
{str2,str3}. Furthermore, it has been shown\cite{dm-nd} that when the
scalar field is massless, the cosmological evolution drives the scalar field
to a value that minimizes the coupling function, towards making the 
contribution of the scalar field very small. This effect makes the theory 
practically indistinguishable from GR at the present epoch.

Solar system tests can put limits on the Post-Newtonian parameters $\gamma$ 
and $\beta$\cite{will1}, which in ST gravity translate into limits on
$a_1$ and $a_2$: the first two coefficients of the Taylor series expansion 
about today of $\alpha (\f)$, the coupling function of the scalar field $\f$ to
matter. We have the relations\cite{ndt,dm-ef}:
\begin{equation}
|\gamma-1|\approx 2a_{1}^{2},\ \beta-1\approx \frac{1}{2}a_{1}^{2}a_2.
\label{eq:gambet}
\end{equation}
From analysis of the Shapiro time delay in the Viking
Mars data\cite{vm}, and from VLBI radio wave deflection\cite{vlbi} we have
\begin{equation}
a_{1}^{2}<1\times10^{-3}, \label{eq:ppngam}
\end{equation}
while Lunar Laser Ranging\cite{llr} gives the limit
\begin{equation}
|1+a_2|a_{1}^{2}<5\times10^{-4}. \label{eq:ppnbet}
\end{equation}
These limits (the most restrictive known to the authors) imply that at the
present epoch in the weak field regime gravity's behavior does not differ much
from that predicted by GR. We will show that the cosmological constraint places
stronger limits on $a_1$ if $a_2$ is positive. In addition it implies that if 
there is a scalar field contribution to gravity, it must be very small from 
the epochs of $e^+e^-$ annihilation and nucleosynthesis up to the present.
The cosmological constraint has been analyzed less generally by Casas \emph{et.
al.}\cite{cgq}, Kalligas \emph{et. al.}\cite{knw}, and Serna and Alimi\cite
{sr-al}. A preliminary version of our results appears in~\cite{wag-ka}.
 
Throughout this paper we follow the convetions of MTW\cite{mtw}($G_{o}=c=1$). 

\section{Scalar-Tensor Theories of Gravity}
\label{sec-ST}

We consider a theory of gravitation in which the gravitational interaction is
carried by the ususal spin-2 field, the metric $\g$, plus a spin-0 field, a 
scalar field $\f$. We work in the Einstein conformal frame of
\cite{dm-ef,dm-nd}, which we call the spin-frame, because the spin-2 and spin-0
excitations are then decoupled from each other. The main advantage of the 
spin-frame is that, in general, calculations of the behavior of the fields (as
opposed to the motion of matter) are more readily manageable. Another advantage
of the spin-frame is that the Cauchy problem is well posed\cite{dm-ef}.

The field equations for our theory follow from extremizing the action
\begin{equation}
S=\frac{1}{16\pi}\int d^{4}x\sqrt{-g}\,[R - 2\gu\partial_{\mu}\f\partial_{\nu}
\f]\;\; +\;\; S_{m}. \label{eq:S}
\end{equation}
Here $g\equiv \det{\g}$, $R\equiv \gu R_{\mu \nu}$ is the Ricci scalar, and
$S_{m}$ is the matter action, which includes all the matter fields and their
interactions including those with the gravitational fields. The matter fields
couple to the gravitational fields through the physical metric
\begin{equation}
\tilde{g}_{\mu \nu}=A^{2}(\f)\g. \label{eq:physg}
\end{equation}
From the previous equation we see that the scalar field couples conformally to
matter via the function $A(\f)$. A priori $A(\f)$ is a completely unspecified
function of the scalar field. Solar system experiments constrain this function
somewhat, and we will show that cosmology can give additional strong 
restrictions on this function. 

The three functions
\begin{eqnarray}
a(\f)\equiv \ln A(\f) \label{eq:a} \\
\alpha(\f)\equiv \frac{\partial a(\f)}{\partial \f} \label{eq:alf} \\
\kappa(\f)\equiv \frac{\partial \alpha(\f)}{\partial \f}\, ,
\end{eqnarray}
will be very useful because they are related to observable quantities. It will
be convenient to express these functions as their Taylor series expansion about
the asymptotic cosmological value of the scalar field in the present epoch, 
$\f_{o}$:
\begin{eqnarray}
a(\f)=a_{1}(\f - \f_{o}) + \frac{a_{2}}{2}(\f - \f_{o})^{2} + \cdots \\
\alpha(\f)=a_{1} + a_{2}(\f - \f_{o}) + \cdots \\
\kappa(\f)=a_{2} + \cdots\, ,
\end{eqnarray}
where we scale our coordinates so that $a(\f_{o})=0$.

When we vary the action with respect to $\gu$ and $\f$, we obtain the field
equations
\begin{eqnarray}
\G=8\pi T_{\mu \nu} + 2\,(\f_{,\mu} \f_{, \nu} -\frac{1}{2}\g\f^{,\sigma}
\f_{,\sigma}) \label{eq:G} \\
\Box \f = -4\pi \alpha(\f)T\, , \label{eq:f}
\end{eqnarray}
where
\begin{eqnarray}
T_{\mu \nu}=\frac{2}{\sqrt{-g}}\frac{\delta S_{m}}{\delta \gu} \label{eq:T}\\
T=\gu T_{\mu \nu} \\
\Box \f= \f^{, \mu}\,_{;\mu}
\end{eqnarray}
are the spin-frame stress-energy tensor, its trace, and the divergence of 
the gradient of $\f$. The Bianchi identity leads to the equations of motion
\begin{equation}
T_{\mu}\,^{\nu}\,_{; \nu}= \alpha(\f)T \f_{, \mu}\,. \label{eq:bianchi}
\end{equation}
In all these formulae we follow the usual convention that comas represent
partial derivatives and semi-colons represent covariant derivatives. 

From the field equations we see that if we set $\f=K,\ \g=g^{(GR)}_{\mu \nu}$
with $K$ a constant such that $\alpha(K)=0$ and $g^{(GR)}_{\mu \nu}$ a solution
to Einstein's equations of GR, we have a solution to our field equations. It 
follows that all general relativistic solutions are also solutions in ST 
gravity with a appropriate choice of constant scalar field. Moreover since 
$\alpha(\f) \equiv \frac{\partial a}{\partial \f}$, $K$ locates an extremum of 
our coupling function $a(\f)$. When $K$ produces a minimum of $a(\f)$, the GR 
solutions are stable solutions of ST gravity, while if $K$ produces a maximum 
or a inflection point the solutions are unstable. It has been demonstrated\cite
{dm-nd} that if $a(\f)$ has a minimum or approaches one symptotycally, then the
cosmological evolution of during the matter era drives the scalar field towards
the minimum value of $a$, and therefore the universe 
evolves towards general relativity.

We can express\cite{dm-ef,will1} the measured gravitational constant far from
all matter,$\ \tilde{G}$, and the PPN parameters $\gamma$ and $\beta$ in 
terms of the Taylor expansion coefficients of $\ a(\f)$
\begin{eqnarray}
\tilde{G}=1 + a_{1}^{2} \\
\gamma-1=\frac{-2a_{1}^{2}}{1+a_{1}^{2}} \\
\beta-1=\frac{1}{2}\frac{a_{1}^{2}a_{2}}{(1+a_{1}^{2})^2}\, .
\end{eqnarray}
From these last formulae we see how solar system tests give us information 
about the coupling function $A(\f)$, i.e., by constraining the possible values 
of $a_{1}$ and $a_{2}$. Our cosmological constraint will give us information
on these same two Taylor series coefficients which, in turn, can be translated
into a restriction on $\gamma$.

\section{Scalar-Tensor Cosmologies}
\label{sec-cosm}

We now turn to the study of the evolution of the universe as a whole in ST 
gravity. We assume the universe is homogeneous and isotropic, so that the
spacetime has a Friedmann-Robertson-Walker metric
\begin{equation}
ds^{2}=-dt^{2}+R^{2}(t)\,[\frac{dr^{2}}{1-kr^{2}}+r^{2}d\Omega ^{2}]\, ,
\end{equation}
with $d\Omega$ being the element of solid angle and $k=-1,\ 0,\ \mbox{or}\ 1$ 
according to whether the universe is open, flat, or closed. In order to compare
with observations we need to express all the physically meaurable quantities
(proper time, expansion rate, density, pressure, etc.) in terms of  spin-frame
quantities (the ones we use for our analysis). Because of the relation 
(\ref{eq:physg}) between the physical metric and the spin-frame metric, the 
proper time measured by ideal clocks is not $d\tau=\sqrt{-ds^{2}}$; it is
\begin{equation}
d\tilde{\tau}=A(\f)\sqrt{-ds^{2}}\, .
\end{equation}
In particular, for comoving observers $d\tilde{\tau}=A(\f)dt$. We shall call 
$\tau$ the spin-frame ``proper'' time, but rememeber that it is not the time 
measured by ideal comoving clocks. The physical scale factor of the universe is
\begin{equation}
\tilde{R}(t)=A(\f)R(t)
\end{equation}
instead of $R(t)$, the spin-frame scale factor. The condition of homogeneity 
implies that our scalar field is only a function of the the spin-frame time 
coordinate, $\f=\f(t)$.

The matter in the universe behaves to a large degree like a perfect fluid
\begin{equation}
T^{\mu \nu}=(\rho + p)u^{\mu}u^{\nu}+p\gu \, .
\end{equation}
In the previous equation 
\begin{equation}
u^{\mu}\equiv \frac{dx^{\mu}}{d\tau}
\end{equation}
is the spin-frame 4-velovity normalized to $\g u^{\mu}u^{\nu}=-1$ for massive
particles. Since $d\tilde{\tau}=A(\f)d\tau$, we have $u^{\mu}=A(\f)\tilde{u}^
{\mu}$ with 
\begin{equation}
\tilde{u}^{\mu}\equiv \frac{dx^{\mu}}{d\tilde{\tau}}
\end{equation}
being the physical 4-velocity, i.e., the actual 4-velocity particles have. The 
physical stress-energy tensor is defined by
\begin{equation}
\tilde{T}_{\mu \nu}=\frac{2}{\sqrt{-\tilde{g}}}\frac{\delta S_{m}}
{\delta \tilde{g}^{\mu \nu}}\, , 
\end{equation}
analogous to equation (\ref{eq:T}) for the spin-frame stress-energy tensor. 
Using the relation (\ref{eq:physg}) between the physical metric and the 
spin-frame metric, we get 
\begin{equation}
T_{\mu \nu}=A^{2}(\f)\tilde{T}_{\mu \nu}\, ,
\end{equation}
which leads us to the following expressions for the spin-frame density and 
pressure of the matter in terms of the physical quantities,
\begin{equation}
\rho=A^{4}(\f)\tilde{\rho},\ p=A^{4}(\f)\tilde{p}\, . \label{eq:p}
\end{equation}
We remind readers that quantities with a tilde are the physical ones, which
are the usual measured values.

Substituting the metric for our isotropic and homogeneous cosmological model 
and the perfect fluid spin-frame stress-energy tensor into our field equations 
(\ref{eq:G}) and (\ref{eq:f}), we obtain
\begin{eqnarray}
-3\frac{\ddot{R}}{R}=4\pi (\rho + 3p) + 2(\dot{\f})^{2}\,, \label{eq:acc} \\
3\left( \frac{\dot{R}}{R} \right)^{2} + 3\frac{k}{R^{2}}=8\pi \rho + (\dot{\f})
^{2} \label{eq:h} \\
\ddot{\f} + 3\frac{\dot{R}}{R}\dot{\f}=-4\pi \alpha (\f)(\rho - 3p) \, , 
\label{eq:sc}
\end{eqnarray}
where the overdots denote derivatives with respect to spin-frame coordinate
time, $t$. From the Bianchi identity (\ref{eq:bianchi}) we obtain the 
``conservation'' equation
\begin{equation}
d(\rho R^{3}) + pd(R^{3})=(\rho - 3p)R^{3}da(\f)\, , \label{eq:cons}
\end{equation}
which, of course, is a consequence of the field equations.

Just as is the case in GR, in ST gravity we need an equation of state in order
to solve our field equations. In general, equations of state relating density 
and pressure are valid only for physical quantities, $\tilde{p}=\tilde{p}(
\tilde{\rho})$, and not spin-frame quantities. All the cosmological epochs we 
consider, except the $e^{+}e^{-}$ era, have simple equations of state of the
type $\tilde{p} \propto \tilde{\rho}$. We will have a similar type of relation
between the spin-frame quantities because of equation (\ref{eq:p}) and the 
simple form of the equation of state. The $e^{+}e^{-}$ era is handled by 
applying first order perturbation theory in the effect of the scalar field 
(arguing why it is not necessary to go to higher orders). To first order the 
spin-frame density and pressure are equal to the physical ones, enabling us to
use the same equation of state for the spin-frame quantities.

\section{Cosmological Evolution and Nucleosynthesis}
\label{sec-cosmnuc}

We now turn our attention to the question of what cosmology says about ST 
gravity. As expected, because of the presence of the scalar field, the
expansion rate of the universe will be different than the one predicted from 
GR. Nucleosynthesis is sensitive to the speed-up factor $\xi_{n}$ (the ratio of
the expansion rate at nucleosynthesis to that predicted by GR). Thus we see 
that the production of primordial light elements, especially $\ ^{4}$He, will 
provide us with limits on ST gravity. In order to obtain these constraints we 
will integrate semi-analytically (in perturbation theory) our cosmological 
equations from the present era back to the epoch of $e^{+}e^{-}$ annihilation.
This last era is handled numerically, performing the appropriate matching at 
the transitions between eras. 

The next step is to establish the relation between the speed-up factor, the
scalar field, and the other quantities of our theory. The following relation 
between the speed-up factor, $\xi_{n}$, and the conformal factor, $A(\f)$, has
been derived~\cite{dm-nd} from quite general considerations,
\begin{equation}
\xi_{n} = \frac{A(\f_{n})}{\sqrt{1+\alpha^{2}(\f_{o})}}\, , \label{eq:xi}
\end{equation}
with $\f_{n}$ being the value of the scalar field at the time of 
nucleosynthesis and $\f_{o}$ its value today. Since $\xi_{n}$ cannot differ 
from unity by more than $\sim 15$\% without over- or under producing$\ ^{4}$He 
~\cite{wag2,wag3} we can Taylor expand our coupling function about $\f_{o}$ and
consider terms beyond the second order as unimportant for all times between
the present and nucleosynthesis:
\begin{equation}
a(\f)=a_{1}(\f - \f_{o}) + \frac{a_{2}}{2}(\f - \f_{o})^{2}\, . \label{eq:aexp}
\end{equation}
Combining the two previous equations with equation (\ref{eq:a}) we arrive at
\begin{equation}
\ln \xi_{n} =a_{1}(\f_{n} - \f_{o})  + \frac{a_{2}}{2}(\f_{n} - \f_{o})^{2} -
\frac{1}{2}a_{1}^{2}
\, , \label{eq:lxi}
\end{equation}
to the desired order.

The equations of motion can be integrated to yield $\f_{n} - \f_{o}$. 
When this quantity is substituted in our last equation we will have the 
speed-up factor as a function of the the Taylor coefficients $a_{1}$ and $a_{2}
$. Then we can use the abundances of primordial light elements ( functions of 
$\xi_{n}$) to obtain the allowed values of $a_{1}$ as a function of $a_{2}$. 

We will restrict ourselves to the study of those ST cosmologies that evolve
towards GR. This is equivalent to considering only coupling functions $a(\f)$
that at least have a minimum or approach one asymptotically. In terms of the 
Taylor expansion coefficients of the coupling function, this implies $a_{2}>0$.
Excluded is the Brans-Dicke case [$a(\f)$ linear in $\f$], which is not that
interesting because the deviations from GR are fixed, i.e., $\alpha$ does not 
depend on the scalar field. Excluded also are the cases where the coupling 
function has no minimum, has points of inflection and at most one maximum. In 
this case solar system constraints place us very near a point of inflection or 
the maximum. In either case the coupling function will evolve away from such a 
point and, in the future, the deviations from general relativity will be more 
pronounced. We will consider both flat and open universes. For each universe 
we will have the following eras: $e^+e^-$ annihilation, radiation dominated, 
and matter dominated. For the open universe we also have a curvature dominated
epoch. We approximate the transitions between eras as sudden.

\subsection{Scalar Field Perturbation Theory}
\label{ssec-per}

Scalar-tensor theories are very general because the coupling function $a(\f)$ 
could be any function of the scalar field. This generality prevents one from 
performing any calculation unless one chooses a particular coupling function.
Thus it seems that any calculations will be dependent on the choice among the 
uncountably many coupling functions. This appears quite unnatural especially 
since the coupling function $a(\f)$ should be a fundamental property of nature
($a$ constant corresponding to GR). Therefore $a(\f)$ should be derived from 
fundamental principles or somehow be obtained from experimental facts. We try 
to do the latter in the present work using the light element abundances.

The function $\alpha(\f)= \frac{da}{d\f}$ acts as a measure of the deviations 
from GR. When $|\alpha| \ll 1$ the scalar contribution to gravity is small and 
the theory is very close to GR, while if $\alpha \gtrsim 1$ there will be large
deviations from GR. The solar systems limits on the PPN parameters $\gamma$ and
$\beta$ mentioned in equations (\ref{eq:ppngam}) and (\ref{eq:ppnbet}) imply 
that the value of $\alpha$ today, $a_{1}$, is small. So at the present time the
universe has small deviations from a general relativistic behavior. This 
suggests that we can treat the effects of the scalar contribution as a first 
order perturbation on GR.  Since we are very near a minimum of $a(\f)$, $a_{1}
$ acts as a ``measure'' of how far we are from that minimum, i.e., the strength
of the scalar contribution to gravity. For $a_{1}=0$, we have no scalar 
contribution (i.e., GR) while for $|a_{1}| \ll 1$ we have small scalar 
contributions. At present, since the effect of the scalar field is small, we 
expect $|a_{1}| \ll 1$  and the cosmological field today, $\f_{o}$, to be 
different by order $a_{1}$ from a value that minimizes $a(\f)$, i.e., $\alpha
(\f)=0$. Moreover, as long as we can apply first order perturbation theory we 
expect the deviations of the scalar field from its value today to also be
proportional to $a_{1}$:
\begin{equation}
\f=\f_{o} + a_{1} \f_{1} + O(a_{1}^{2})\, .
\end{equation}
Therefore we can consider $a_{1}$ as the small parameter for a first order 
perturbation theory expansion about general relativity. Besides $\f$ itself, 
$\alpha$ is the only other scalar field dependent quantity to appear in the 
field equations (\ref{eq:G}) and (\ref{eq:f}). Hence to do first order 
perturbation theory in the effect of the scalar field we need $\alpha$ to 
linear order in $\f$:
\begin{equation}
\alpha(\f)=a_{1} + a_{2}(\f - \f_{o}) + O(2)=a_{1}(1+a_{2}\f_{1})+O(a_{1}^{2})
\, .
\end{equation}
Since this is the derivative with respect to $\f$ of equation (\ref{eq:aexp}), 
we see that we need $a(\f) \mbox{, or equivalently}\ A(\f)$, to second order. 
But if higher order deviations are important, we will over- or underproduce 
$\ ^{4}$He. Therefore our first order perturbation theory approach is valid at 
least up  to the time of nucleosynthesis. If we look at the field equations 
(\ref{eq:G}) for the metric, we see that they depend on derivatives of $\f$ 
squared. Since the physical metric as given by equation (\ref{eq:physg}) 
differs from the spin-frame metric by quantities of second order ($a_{1}^{2}$),
it must differ from the GR metric by quantities of at least second order.
Moreover, the matter lagrangian, and hence, the stress-energy tensor, will be 
equal to the GR quantities to at least zeroth and first order. Therefore the 
metric will be unaffected to first order while to zeroth order we just have a 
GR solution, 
$g^{(GR)}_{\mu \nu}$: 
\begin{equation}
\g = g^{(GR)}_{\mu \nu} + O(a_{1}^{2}) \, .
\end{equation}
In general, the physical metric is of more interest. The first order correction
in the scalar field affects the physical metric to second order. The second 
order correction to the metric affects ideal clocks and measuring sticks to the
same order as the first order scalar field correction. Therefore we would need 
to find this metric correction to understand the full physics of a problem to 
the desired order. For the nucleosynthesis constraint we are fortunate because 
we have formula (\ref{eq:xi}) or (\ref{eq:lxi}) which relates the speed-up 
factor to the coupling function independently of the metric. We see that in 
this particular case we can neglect the back reaction of the metric in the 
presence of the scalar field. There will also be corrections to the spin-frame 
stress-energy tensor, but as we previously argued they will be second order:
\begin{equation}
T_{\mu \nu}=T^{(GR)}_{\mu \nu} + O(a^{2}_{1})\, ,
\end{equation}
with $T^{(GR)}_{\mu \nu}$ the general relativistic stress-energy tensor. To 
this order the spin-frame stress-energy tensor is equal to the physical one.

Now we consider the ST cosmological and ``conservation'' equations (\ref
{eq:acc}-\ref{eq:cons}) in perturbation theory. We note that these equations 
are not independent (they are related through the Bianchi identity). Hence we 
do not use equation (\ref{eq:acc}). For the spin-frame expansion factor, scalar
field, density, and pressure we have the following relations to the desired 
order:
\begin{eqnarray}
R(t) = R^{(GR)}(t) + O(a_{1}^{2}) \\
\f=\f_{o} + a_{1}\f_{1} + O(a_{1}^{2}) \label{eq:exphi}\\
\rho(t)=\rho^{(GR)}(t) + O(a_{1}^{2}) \\
p(t)=p^{(GR)}(t) + O(a_{1}^{2})\, .
\end{eqnarray}
When we substitute these in the cosmological equations we obtain to zeroth
(GR) order:
\begin{eqnarray}
3\left( \frac{\dot{R}}{R} \right)^{2} + 3\frac{k}{R^{2}}=8\pi \rho 
\label{eq:h0}\\
d(\rho R^{3}) + pd(R^{3})=0 \label{eq:cons0}\, ,
\end{eqnarray}
where we have dropped the $(GR)$ superscripts.
To first order scalar field equation becomes
\begin{equation}
\ddot{\f_{1}} + 3\frac{\dot{R}}{R}\dot{\f_{1}}
+4 \pi (1 + a_{2}\f_{1})(\rho - 3p) =0\, .
\end{equation}
For later use we define
\begin{equation}
\psi \equiv 1 + a_{2}\f_{1} \, ,\label{eq:psi}
\end{equation}
so that $\psi$ satisfies
\begin{equation}
\ddot{\psi} + 3\frac{\dot{R}}{R}\dot{\psi} +
4 \pi a_{2}(\rho - 3p)\psi=0 \label{eq:dpsi}\, .
\end{equation}
For the nucleosynthesis constraint we will need the speed-up factor. 
Substituting the perturbation expansion (\ref{eq:exphi}) of $\f$ into the 
relation (\ref{eq:lxi}) for the speed-up factor we get
\begin{equation}
\ln \xi_{n} =a_{1}^{2}[-\frac{1}{2} +\f_{1}(t_{n}) + \frac{a_{2}}{2}\f_{1}^{2}
(t_{n})] =a_{1}^{2}(-\frac{1}{2} + \frac{\psi ^{2}(t_{n})-1}{2a_{2}} ) \, , 
\label{eq:lxiexp}
\end{equation}
where $t_{n}$ is the time at nucleosynthesis, and we used equation (\ref
{eq:psi}) to get the last equality. 

Next we find solutions to the field equations in the different epochs that 
occur in the two types of universe we consider: flat and open.

\subsection{Curvature Dominated Epoch}
\label{ssec-cd}

We consider an era in which the evolution of the universe is mainly driven by 
the curvature term, i.e., the curvture term is large compared to the density
term. This describes to a fairly good aproximation the present epoch for a 
sufficiently open universe ($k=-1$).

When we neglect the density term in our zeroth order field equation (\ref
{eq:h0}) we have
\begin{equation}
\left( \frac{\dot{R}}{R} \right)^{2} - \frac{1}{R^{2}}=0 \, ,
\end{equation}
which trivially integrates to
\begin{equation}
R(t)=t+C_{1} \, , \label{eq:Rcd}
\end{equation}
with $C_{1}$ a constant of integration.

We must consider how the matter density depends on the scale factor since
we should include the matter density term in equation (\ref{eq:dpsi}) for the
scalar field because although it is small at each time it builds up during the
evolution. Taking $p=0$ in the zeroth order conservation equation 
(\ref{eq:cons0}) gives us the well known simple result
\begin{equation}
\rho_{m}=\frac{M_{o}}{R^{3}} \, , \label{eq:rhom}
\end{equation}
where $M_{o}$ is a constant of normalization and the subscript $m$ on $\rho$ is
to specify that it is the matter density.

To first order we use the auxiliary scalar field $\psi$ and its corresponding 
equation (\ref{eq:dpsi}). Using our last result the $\psi$ equation becomes
\begin{equation}
\ddot{\psi} + \frac{3}{t+C_{1}} \dot{\psi} + \frac{4 \pi a_{2} M_{o}}{(t+C_{1})
^{3}} \psi=0 \, .
\end{equation}
This is a transformed Bessel's equation\cite{math}, which has solution
\begin{equation}
\psi(t)=\frac{C_{2}}{t+C_{1}}J_{2} \left( \sqrt{\frac{16 \pi M_{o} a_{2}}{t+
C_{1}}} \right) + \frac{C_{3}}{t+C_{1}}Y_{2} \left(\sqrt{\frac{16 \pi M_{o} a_
{2}}{t+C_{1}}} \right) \, ,  \label{eq:psicd}
\end{equation}
where the $C_{k}$ are constants of integration, and $J_{2}(x)$ and $Y_{2}(x)$
are Bessel functions of the first and second kind respectively. 

It is interesting that $\psi$ in equation (\ref{eq:psicd}) does not converge
to zero for large times, rather it approaches a constant. To first order $\psi
=0$ corresponds to the value of $\f$ for which $\alpha(\f)=0$. Therefore the
universe does not evolve towards a state in which the deviations from GR 
disappear, but as it evolves the deviation from GR converges to a finite,
though small, value. The attractor mechanism of \cite{dm-nd} is not too 
effective because for large times the Hubble damping makes the scalar field 
``stop'' before reaching the value which produces $\alpha=0$. The full 
attractor mechanism is present only for flat scalar-tensor cosmologies because 
those will have GR as a limit for large times.

\subsection{Matter Dominated Epoch}
\label{ssec-md}

In a matter dominated epoch the evolution of the universe is driven by the 
density of the matter (described as a fluid of nonrelativistic particles). 
Hence the matter density is much larger than either the curvature term or the 
density and pressure of radiation. Such an epoch corresponds to the present for
a flat universe ($k=0$), or to the epoch previous to the curvature dominated 
one in an open universe ($k=-1$).

The zeroth order field equation (\ref{eq:h0}) becomes
\begin{equation}
3\left( \frac{\dot{R}}{R} \right)^{2}=8 \pi \frac{M_{o}}{R^{3}}\, .
\end{equation}
This equation is easily integrated to yield
\begin{equation}
R(t)=(6 \pi M_{o})^{\frac{1}{3}}(t + D_{1})^{\frac{2}{3}} \, , \label{eq:Rmd}
\end{equation}
with $D_{1}$ a constant of integration.

We need the scalar field evolution to first order. Just as we did in the
previous section (\ref{ssec-cd}), we use the auxiliary field $\psi$. When
the results (\ref{eq:rhom}) and (\ref{eq:Rmd}) are substituted in the $\psi$ 
equation (\ref{eq:dpsi}) we have
\begin{equation}
\ddot{\psi} + \frac{2}{t + D_{1}} \dot{\psi} + 
\frac{2a_{2}}{3(t + D_{1})^{2}} \psi =0 \, .
\end{equation}
By making the variable change $x=\ln (t + D_{1})$ we end up with an equation 
for a damped harmonic oscillator:
\begin{equation}
\psi'' + \psi' + \frac{2}{3}a_{2} \psi =0 \, ,
\end{equation}
where primes denote derivatives with respect to $x$. This equation has three
types of solution (under, over, or critically-damped), according to the value 
of $a_{2}$. \newline
\underline{Case 1} (underdamped: $a_{2}>\frac{3}{8}$):
\begin{equation}
\psi (t)= \frac{D_{2}}{\sqrt{t + D_{1}}} \sin\left[ \sqrt{\frac{2}{3}a_{2} - 
\frac{1}{4}} \ln\left( \frac{t + D_{1}}{D_{3}} \right) \right] \, . \label
{eq:psimdu}
\end{equation}
\underline{Case 2} (criticallydamped: $a_{2}=\frac{3}{8}$):
\begin{equation}
\psi (t)= \frac{D_{2}}{\sqrt{t + D_{1}}} \ln\left( \frac{t + D_{1}}{D_{3}} 
\right) \, . \label{eq:psimdc}
\end{equation}
\underline{Case 3} (overdamped: $a_{2}<\frac{3}{8}$):
\begin{equation}
\psi (t) = \frac{D_{2}}{\sqrt{t + D_{1}}} \sinh\left[ \sqrt{\frac{1}{4}- \frac
{2}{3}a_{2}} \ln\left( \frac{t + D_{1}}{D_{3}} \right) \right] \, . \label
{eq:psimdo}
\end{equation}
In all three cases $D_{k}\,(k=2 \mbox{ and } 3)$ are constants of integration.

\subsection{Radiation Dominated Epoch}
\label{ssec-rd}

Previous to the matter era, there is an epoch in which the density and pressure
of relativistic particles (radiation) is large compared to the density of the 
matter (nonrelativistic fluid) as well as the curvature term. This epoch 
describes the universe fairly well prior to matter-radiation equality.

Remembering that the equation of state for radiation is $p_{r}=\rho_{r}/3$, our
zeroth order conservation equation (\ref{eq:cons0}) becomes
\begin{equation}
0=d(\rho_{r} R^{3}) + \frac{\rho_{r}}{3}d(R^{3}) = \frac{d(\rho_{r} R^{4})}{R} 
\, ,
\end{equation}
which trivially gives
\begin{equation}
\rho_{r}= \frac{N_{o}}{R^{4}} \, ,\ p_{r}= \frac{N_{o}}{3R^{4}}\, , 
\label{eq:rhor}
\end{equation}
with $N_{o}$ a integration constant. The subscript $r$ specifies that these are
the density and pressure of radiation.

Taking into account this last result, the expansion factor equation (\ref
{eq:h0}) to zeroth order is
\begin{equation}
3\left( \frac{\dot{R}}{R} \right)^{2}=8 \pi \frac{N_{o}}{R^{4}}\, .
\end{equation}
We solve this equation to find
\begin{equation}
R(t)= \left(\frac{32 \pi N_{o}}{3} \right)^{\frac{1}{4}} (t + B_{1})^{\frac{1}
{2}} \, . \label{eq:Rrd}
\end{equation}

To first order we must solve equation (\ref{eq:dpsi}) for $\psi$. The last
term in this equation is proportional to $\rho - 3p$, which vanishes if we
consider radiation. On the other hand the matter density ($\rho_{m}=\frac
{M_{o}}{R^{3}}$) also contributes to the total density. This is a small 
correction at any particular time in the radiation epoch, but it builds up as 
we integrate the $\psi$ equation over the whole radiation era. Therefore we 
must include the effect of the matter density in the scalar field equation. We 
find
\begin{equation}
\ddot{\psi} +\frac{3}{2(t + B_{1})}\dot{\psi} + \frac{a_{2}\chi}{(t + B_{1})^
{\frac{3}{2}}} \psi=0 \, ,
\end{equation}
where we have introduced
\begin{equation}
\chi \equiv \frac{3M_{o}}{4N_{o}} \left(\frac{2 \pi N_{o}}{3} \right)^{\frac{1}
{4}} \, . \label{eq:chi}
\end{equation}
This is a transformed Bessel's equation\cite{math}, which has solution
\begin{equation}
\psi(t)=\frac{B_{2}}{(t + B_{1})^{\frac{1}{2}}}J_{1}[4(a_{2} \chi)^{\frac{1}
{2}} (t + B_{1})^{\frac{1}{4}}] +\frac{B_{3}}{(t + B_{1})^{\frac{1}{2}}}Y_{1}
[4(a_{2} \chi)^{\frac{1}{2}} (t + B_{1})^{\frac{1}{4}}] \, , \label{eq:psirint}
\end{equation}
where $J_{1}(x)$ is a Bessel function of the first kind and $Y_{1}(x)$ is a 
Bessel function of the second kind. In addition $B_{k}\,(k=2 \mbox{ and } 3)$ 
are constants of integration. Without loss of generality we can set $B_{3}=0$.
At the time when we match the radiation era solution for $\psi$ to the matter 
epoch, the argument of the Bessel functions in equation (\ref{eq:psirint}) is 
large. For $x \gg 1$ we have $Y_{1}(x) \ll J_{1}(x)$ so setting $B_{3}=0$ is 
consistent at the time of matching. To see whether our assumption is correct we
note that the radiation era starts at a time a lot smaller that when it ends. 
Therefore the matching at the other end might be troublesome because for $x \ll
 1$ we have $Y_{1}(x) \gg J_{1}(x)$. To see that we do not run into problems we
first let $x=4(a_{2} \chi)^{\frac{1}{2}} (t + B_{1})^{\frac{1}{4}}$ so that 
(redefining $B_{k}\, ,\ k=2 \mbox{ and } 3\, ,$ by multiplication by an 
appropriate constant) we get
\begin{equation}
\psi(x)=\frac{B_{2}}{x^{2}}J_{1}(x) +\frac{B_{3}}{x^{2}}Y_{1}(x) \, . 
\end{equation}
If we let $\psi_{*} \ \mbox{and}\ \psi'_{*}$ be the values of $\psi(x_{*}) \ 
\mbox{and}\ \frac{d\psi}{dx}(x_{*})$ at the point of matching (call it $x_{*}$)
to the epoch previous to the radiation era, we have
\begin{equation}
\frac{B_{3}}{B_{2}} = -\frac{\psi_{*}J_{2}(x_{*}) + \psi'_{*}J_{1}(x_{*})}{
\psi_{*}Y_{2}(x_{*}) +\psi'_{*}Y_{1}(x_{*})} \, .
\end{equation}
Therefore even if the matching is done at $x_{*} \ll 1$ we have
\begin{equation}
 \left| \frac{B_{3}}{B_{2}} \right| \sim O(x_{*}^{3}) \ll 1 
\end{equation}
at worst. Hence we are justified in taking $B_{3}=0$ throughout the radiation
epoch. Finally we obtain
\begin{equation}
\psi(t)=\frac{B_{2}}{(t + B_{1})^{\frac{1}{2}}}J_{1}[4(a_{2} \chi)^{\frac{1}
{2}} (t + B_{1})^{\frac{1}{4}}]  \, . \label{eq:psird}
\end{equation}

\subsection{Epoch of $e^{+}e^{-}$ Annihilation}
\label{ssec-e+e-}

The last era we consider is the one when the electron-positron pairs annhilate.
During the early evolution of the universe, as we have mentioned, the 
cosmological fluid can be approximated as a perfect fluid with a radiation 
equation of state. Consider some species $i$ of particles and antiparticles 
with mass $m_{i}$. When the temperature of the universe passes through the 
threshold $kT \sim m_{i}c^{2}$, the pairs of $i$ particles and antiparticles 
annihilate, creating important though small deviations from the radiation 
equation of state. These deviations act as a source for our scalar field, 
giving it an extra kick. In particular, we are interested in the thershold for 
$e^+ e^-$ annihilation, $T \simeq 5 \times 10^{9}$ K, because it is the only
annihilation that occurs after the neutron-proton ratio has frozen out of 
equilibrium, occuring around the same time that we have synthesis of primordial
elements. 

We will start the $e^+ e^-$ epoch at the time of neutrino decoupling, $T \simeq
3 \times 10^{10}$ K. Around such time the universe can be safely described as 
radiation including the behavior of the $e^+ e^-$ pairs because they are still
relativistic. We will terminate it when the temperature drops to $T\simeq 3 
\times 10^{8}$. At this point we can once again approximate the cosmological 
fluid as radiation since, even though the leftover electrons are not 
relativistic their density and pressure are very small. Therefore we match with
the radiation era solution at this time.

As derived in Appendix \ref{app-thrm} we have the following densities and 
pressures as functions of temperature:
\begin{eqnarray}
\rho_{\gamma}(T)= \frac{\pi ^{2} (kT)^{4}}{15 \hbar^{3}} \label{eq:erhogam} \\
p_{\gamma}(T)= \frac{\rho_{\gamma}(T)}{3} \label{eq:epgam} \\
\rho_{\nu}(T_{\nu})= \frac{21}{8} \rho_{\gamma}(T_{\nu}) \label{eq:erhonu} \\
p_{\nu}(T_{\nu})= \frac{\rho_{\nu}(T_{\nu})}{3} \label{eq:epnu} \\
p_{e}(T) \simeq \frac{2 m_{e}^{4}}{\pi ^{4}\hbar ^{3}} \times [(\frac{kT}{m_
{e}})^{2} K_{2}(\frac{m_{e}}{kT}) - (\frac{kT}{2m_{e}})^{2}K_{2}(\frac{2m_{e}}
{kT})] \label {eq:epe} \\
\rho_{e}(T) \simeq 3p_{e}(T) + \frac{2 m_{e}^{4}}{\pi ^{4}\hbar ^{3}} \times
[(\frac{kT}{m_{e}})K_{1}(\frac{m_{e}}{kT}) - (\frac{kT}{2m_{e}})K_{1}(\frac{2m_
{e}}{kT})] \label {eq:erhoe} \, .
\end{eqnarray}
Here $\hbar$ is Planck's constant divided by $2 \pi$, $k$ is Boltzmann's 
constant, and $m_{e}$ is the electron rest mass. The subscripts $\gamma \, , \
 \nu \, , \mbox{ and } e$ refer to photons, neutrino-antineutrino pairs, and 
electron-positron pairs respectively. The functions $K_{i}\,(i=1 \mbox{ and } 2
)$ are hyperbolic Bessel functions.

In Appendix \ref{app-thrm} we obtain the following differential equations for
the scale factor and time as functions of temperature:
\begin{eqnarray}
\frac{d}{dT} \left(\frac{R^{3}(T)[\rho_{1}(T) + p_{1}(T)]}{T} \right) = 0
\label{eq:eRT} \\
\frac{dt}{dT} = \frac{\frac{d}{dT} \ln \left[\frac{T}{\rho_{1}(T) + p_{1}(T)} 
\right]}{\sqrt{24 \pi [\rho_{1}(T) + \rho_{\nu}(T)]}} \label{eq:etT} \, ,
\end{eqnarray}
where $\rho_{1}(T)= \rho_{e}(T) + \rho_{\gamma}(T)\, , \ p_{1}(T)= p_{e}(T) + 
p_{\gamma}(T)$.

Writing equation (\ref{eq:dpsi}) for $\psi$ using the temperature as the 
independent variable, we have
\begin{equation}
\frac{d^{2}\psi}{dT^{2}} + f_{1}(T)\frac{d\psi}{dT} + 4\pi a_{2}f_{2}(T)\psi =0
\label{eq:edpsi} \, ,
\end{equation}
where $f_{1}(T) \mbox{ and } f_{2}(T)$ are defined by:
\begin{eqnarray}
f_{1}(T) \equiv \frac{\frac{d}{dT}\left[\frac{dT}{dt} + \ln \left(\frac{T}{
\rho_{1} + p_{1}} \right)\right]}{\frac{dT}{dt}} \\
f_{2}(T) \equiv \frac{\rho_{1} - 3p_{1}}{\left(\frac{dT}{dt} \right)^{2}} \, .
\end{eqnarray}

The $\psi$ equation was integrated numerically using a fourth order Runge-Kutta
method with adaptive stepsize obtained from~\cite{num-rec}. The initial 
conditions for the integration were obtained by matching to the radiation epoch
solution at the lower temperature ($T\simeq 3\times 10^{8}$). There was no need
to solve for the scale factor in the $e^+e^-$ era because the speed-up factor
depends only on the scalar field and everywhere the scale factor appears in the
differential equation for $\psi$, it can be replaced by functions of the 
densities and temperature.

\subsection{Nucleosynthesis}
\label{ssec-nuc}

The production of light elements in the early universe depends on the baryon
density, the number of light($m \lesssim 1$ MeV) neutrinos $N_{\nu}$, the 
neutrino lifetime $\tau_{\nu}$, and the speed-up factor $\xi_{n}$ in such a way
that upper limits on the mass fractions $X( ^{4}\mbox{He})$ and $X(^{2}\mbox{H}
 +\, ^{3}\mbox{He})$ provide an upper limit on $\xi_{n}$. To be more specific, 
the primordial $^{4}$He mass fraction, $X( ^{4}\mbox{He})$, is given 
by\cite{cgq,olv}:
\begin{eqnarray}
X( ^{4}\mbox{He}) = 0.228 + 0.010 \ln ( \eta_{10}) + 0.012(N_{\nu} - 3)+ 
\nonumber \\
0.185 \left( \frac{\tau_{\nu} - 889.8}{889.8}\right) + 0.327 \log\xi_{n} \, ,
\label{eq:xhe}
\end{eqnarray}
with $ \eta_{10} \equiv 10^{10}n(\mbox{baryon})/n(\mbox{photon})$. For the 
primordial mass fraction of $^{2}\mbox{H} +\, ^{3}$He we adopt the conservative
upper limit\cite{olv} $X(^{2}\mbox{H} +\, ^{3}\mbox{He}) < 1.7 \times 10^{-4}$.
This limit implies the inequality\cite{cgq}:
\begin{equation}
\ln ( \eta_{10}) \geq \ln(2.60) + 0.085(N_{\nu} - 3) + 3.7 \log\xi_{n} \, .
\label{eq:eta}
\end{equation}
Combining equations(\ref{eq:xhe}) and (\ref{eq:eta}), we find
\begin{eqnarray}
X( ^{4}\mbox{He})_{\mbox{obs.}} \geq X( ^{4}\mbox{He}) \geq 0.2375 + 0.013(N_
{\nu} - 3)+ 
\nonumber \\
0.185 \left( \frac{\tau_{\nu} - 889.8}{889.8}\right) + 0.364 \log\xi_{n} \, .
\end{eqnarray}
We use the values $N_{\nu} =3$ and $\tau_{\nu} < 885$ sec., which are 
consistent with experimental constraints. We take the reasonable limit $X( ^{4}
\mbox{He}) \leq 0.250$, which implies
\begin{equation}
\ln \xi_{n} \leq 0.0797 \, .\label{eq:xicons}
\end{equation}

\subsection{Flat and Open Universes}
\label{ssec-fltuni}

We now turn our attention to a flat universe (zero spatial curvature, $k=0$). 
For each of these universes we calculated the speed-up factor, which from 
equation (\ref{eq:lxiexp}) we know depends only on the value of the scalar 
field at nucleosynthesis.

In a flat universe, the present epoch can be approximated very well as a matter
dominated era. So our solutions for the scale factor $R(t)$ and the auxiliary
scalar field $\psi$ are the ones given in subsection \ref{ssec-md}. To evaluate
some of the constants of integration we impose the appropriate boundary 
conditions:
\begin{eqnarray}
\frac{\dot{R}}{R}|_{t=t_{o}}= H_{o} = 75\, \mbox{km/sec/Mpc} \\
\psi(t_{o}) = 1 \, .
\end{eqnarray}
These conditions determine two of the constants $D_{k}\,(k=2 \mbox{ and } 3)$ 
in terms of the remaining one. 

Moving back in time we arrive at the radiation dominated epoch, which we 
analyzed in subsection \ref{ssec-rd}. We can take our radiation era scale 
factor, equation (\ref{eq:Rrd}), to describe the behavior of the universe up to
much higher densities, i.e., $B_{1}=0$, because deviations from radiation
dominance (such as annihilation epochs) are short lived and only contribute
very small corrections. Once again to determine the constants of integration we
do the approproate matching at the matter epoch radiation era interface, i.e., 
continuity of $R,\ \psi,\ \mbox{and}\ \dot{\psi}$.

Going farther into the past we get to the era of $e^+e^-$ annihilation. We have
the auxiliary field $\psi$ as a function of $a_{2}$ from the solution in the 
radiation era. Therefore for different values of $a_{2}$, we have different 
values of $\psi$ at nucleosynthesis after the appropriate matching with the 
radiation era solution. Combining equation (\ref{eq:lxiexp}), which expresses 
$\ln \xi_{n}$ in terms of $\psi(t_{n}),\ a_{1},\mbox{ and } a_{2}$, with the 
nucleosynthesis constraint equation (\ref{eq:xicons}) for the speed-up factor 
we obtain allowed values of $|a_{1}|,\mbox{ for a given } a_{2}$ which we plot 
in Figure \ref{fig1} along with the solar system constraint (dotted line). The 
allowed ranges are the regions under the curves. We see from equation 
(\ref{eq:lxiexp}) that for a given $a_{2}$, $|a_{1}| \propto \sqrt{\ln \xi_{n}}
$.

The first thing to note is that throughout most of the range the constraint 
from nucleosynthesis is more restrictive than the solar system constraint. One 
thing that catches the eye are the peaks at $\log(a_{2}) \simeq -0.55\, ,\ 1.74
\, ,\ 2.14\, ,\mbox{ and } 2.43$. Although they appear finite in the graph, 
they are infinite; meaning that at the peaks $a_{1}$ could have any value (no 
constraint at all). This is so because, for these values of $a_{2}$, the 
speed-up factor at nucleosynthesis is unity (the expansion rate at 
nucleosynthesis is the same in GR and in ST gravity with these values of $a_{2}
$). Hence at the peaks the abundances of primordial light elements do not place
any restrictions on $a_{1}$. There will be very small ranges around the
peaks where the nucleosynthesis constraint is less restrictive the solar system
one (the first peak is in the region where the solar system constraint is more 
restrictive anyway). There are  values of $a_{2}$ that produce zeros of $a_{1}
$. If there are any deviations (no matter how small) from GR at these values in
the present, the bounds on primordial light elements will be violated.  Most 
values of $a_{2}$ are consistent with an upper limit of $|a_{1}| \sim 10^{-5}$ 
which implies an upper limit on the PPN parameter $\gamma$ of $|1-\gamma | 
\sim 10^{-10}$.

The analysis of the open universe (negative spatial curvature, $k=-1$) is very 
similar to that of the flat universe. The difference between the two is that 
the present epoch in the open universe is a curvature dominated one, which was 
analyzed in subsection \ref{ssec-cd}. The epochs previous to the present one 
are matter dominated, radiation dominated, and a $e^+e^-$ annihilation era, in 
that order. We take our open universe to have $0.1$ of the critical density, 
i.e., $\Omega_{o}=0.1$. For the open universe we plot in Figure \ref{fig2} the 
$a_{1}$ vs $a_{2}$ constraints. The same discussion as for the flat case
applies here.  We have the upper limits $|a_{1}| \sim 10^{-3}$ and $|1-\gamma |
\sim 10^{-7}$. The main difference with the flat universe results
is that the limits on $a_{1}$ are 1 to 2 orders of magnitude higher, but still
more restrictive than solar system limits over most of the range of $a_{2}$.

\section{Conclusion}
As mentioned in the Introduction, there is motivation to study scalar-tensor 
theories of gravity from the theoretical point of view as well as a means for 
comparison with experimental gravity. 

After noting the difficulty posed by the arbitrariness in the scalar coupling 
function $a(\f)$, we noted that its slope must be small at present in order to 
satisfy solar system limits on PPN parameters. In order to remove the freedom 
in the choice of $a(\f)$, we used perturbation theory about the value of the 
the scalar field today ($\f_{o}$). As long as $\f_{o}$ is close to a value that
minimizes $a(\f)$, the universe will have general relativistic behavior at 
present and peturbation theory will be equivalent to perturbing about GR, 
having the slope of $a(\f)$ as the perturbation parameter. In order to find 
nucleosynthesis constraints on $a_{1}$ and $a_{2}$ we used first order 
perturbation theory because, if higher order correction are important, they 
would lead to violation of present bounds on primordial light elements. 
Moreover, because of equation (\ref{eq:lxi}) for the speed-up factor $\xi 
\simeq 1$ indictes that $\f_{n} \simeq \f_{o}$, we can neglect the back 
reaction on the metric.

We analyzed the scalar field evolution in flat ($\Omega_{o}=0$)and open 
universes ($\Omega_{o}=0.1$) to obtain the speed-up factor at nucleosynthesis 
as a function of $a_{1}$ and $a_{2}$. Using the constraint that the abundances 
of light elements place on $\xi$, allowed values of $a_{1}$ as a function of 
$a_{2}$ were obtained. In both universes the constraint was more restrictive
(for $a_{2} \geq 1/3$) than the one from solar system experiments, the most 
restrictive one coming from a flat universe. This means that if ST gravity is 
the true theory of gravity, it must be practically indistinguishable from GR 
through a significant period of the universe's history at least for these 
cosmologies.

\section*{Acknowledgements}

D. I. Santiago was supported by NASA grant NAS8-39225 to Gravity Probe B. 
D. Kalligas was partially supported by the 1995 PENNED program (no. 512).
Very special thanks to the Friday Gravity sewing circle at GP-B.  

\appendix
\section{Thermodynamics of the Early Universe}
\label{app-thrm}

The purpose of the present appendix is to review and rederive some of the well
known results\cite{alph,wag4} on the densities and pressures of different
species as functions of temperature. Differential equations for the temperature
and the expansion factor are also obtained.

In order to calculate thermodynamic quantities for the different species we 
need the number density of particles in a energy range $dE$. This quantity is
given by
\begin{equation}
n(E)dE=\frac{g_{s}}{2 \pi ^{2} \hbar ^{3}} \times \frac{PEdE}{\exp (\frac{E+
\mu}{kT}) \pm 1} \, ,
\end{equation}
where $g_{s}$ is the number of spin states, $\mu$ is the chemical potential, $P
$ is the momentum, $E$ is the energy, and the $+$($-$) sign corresponds to 
fermions(bosons). For photons and for neutrinos (assuming no neutrino 
degeneracy) we have $\mu = 0$. For electrons and positrons we assume that the 
difference between their numbers is small compared to their sum, so that we
can set $\mu = 0$ for them too (this approximation breaks down when the
density and pressure of electrons and positrons is negligible). 

The densities and pressures of the different species are given by
\begin{eqnarray}
\rho = \int_{m}^{\infty} En(E)dE \\
p = \int_{m}^{\infty} \frac{P^{2}}{E}n(E)dE \, ,
\end{eqnarray}
where m is the mass of the particle species.

For photons and neutrino-antineutrino pairs (assuming three neutrino species) 
we obtain
\begin{eqnarray}
\rho_{\gamma} = \frac{\pi ^{2} (kT)^{4}}{15 \hbar^{3}} \\
p_{\gamma} = \frac{\rho_{\gamma}}{3} \\
\rho_{\nu} = \frac{21}{8}\rho_{\gamma}(T_{\nu}) \\
p_{\nu} = \frac{\rho_{\nu}}{3} \, ,
\end{eqnarray}
where the subscripts $\gamma$ and $\nu$ refer to photons and 
neutrino-antineutrino pairs respectively. For the combined densities of 
electrons and positrons we have
\begin{eqnarray}
p_{e} = \frac{2m_{e}^{4}}{ \pi ^{4} \hbar ^{3}} \sum_{n=1} ^{\infty} (-1)^{n+1}
\left( \frac{kT}{nm_{e}}\right)^{2} K_{2}\left( \frac{nm_{e}}{kT} \right) \\
\rho_{e} = 3p_{e} + \frac{2m_{e}^{4}}{ \pi ^{4} \hbar ^{3}} \sum_{n=1} ^{\infty
} (-1)^{n+1} \left( \frac{kT}{nm_{e}}\right) K_{1}\left( \frac{nm_{e}}{kT} 
\right) \, ,
\end{eqnarray}
where $K_{i}\,(i=1 \mbox{ and } 2)$ are hyperbolic Bessel functions. The 
following will prove useful in some of our next derivations:
\begin{equation}
\frac{dp_{a}}{dT} = \frac{\rho_{a} + p_{a}}{T} \, ,
\end{equation}
with $a=\gamma\, , \ \nu\, , \ \mbox{or}\ e$.

In the epoch of $e^+e^-$ annihilation the neutrinos have just deocoupled, hence
they do not participate in the thermodynamic equilibrium, but are separetely
conserved. The very small baryon contribution is also separetely conserved 
because at these times they have long been decoupled from the equlibrium. 
Photons and $e^+e^-$ pairs interact and are in thermodynamic equilibrium. We
have the conservation equation (\ref{eq:cons0})(which is valid to first order 
in the perturbation expansion of the present paper):
\begin{equation}
\frac{d(\rho R^{3})}{dT} + p \frac{d(R^{3})}{dT} = 0 \, .
\end{equation}
For neutrinos and for baryons (which we take to be pressureless) the 
conservation equation leads to
\begin{eqnarray}
\frac{d(\rho_{\nu} R^{4})}{dT} = 0 \\
\frac{d(\rho_{b} R^{3})}{dT} = 0 \, ,
\end{eqnarray}
with the subscripts $\nu \mbox{ and } b$ refering to neutrinos and baryons
respectevely. For the interacting species, $e^+e^-$ pairs and photons, we
define the following densities and pressures:
\begin{eqnarray}
\rho_{1}= \rho_{e} + \rho_{\gamma} \\
p_{1}= p_{e} = p_{\gamma} \, .
\end{eqnarray}
Substituting these last expressions into the conservation equation and using 
the equality
\begin{equation}
\frac{d\rho_{1}}{dT}= \frac{d(\rho_{1} + p_{1})}{dT} - \frac{dp_{1}}{dT} =
 \frac{d(\rho_{1} + p_{1})}{dT} - \frac{\rho_{1} + p_{1}}{T} \, ,
\end{equation}
we obtain
\begin{equation}
\frac{3}{R} \frac{dR}{dT} = \frac{d}{dT} \ln \left( \frac{T}{\rho_{1} + p_{1}}
\right) \, .
\end{equation}
The last equation can be written in the much more compact form
\begin{equation}
 \frac{d}{dT} \left( \frac{R^{3}[\rho_{1} + p_{1}]}{T}\right) \, .
\end{equation}
If we combine this result with the field equation (\ref{eq:h0}) for the scale
factor we end up with
\begin{equation}
\frac{dt}{dT} = \frac{\frac{d}{dT} \ln \left[\frac{T}{\rho_{1} + p_{1}} 
\right]}{\sqrt{24 \pi \rho}} \, , 
\end{equation}
where
\begin{equation}
\rho = \rho_{1} + \rho_{\nu} + \rho_{b} \simeq \rho_{1} + \rho_{\nu}
\end{equation}
is the total density.

\begin{figure}[h]
\caption{ $\log (|a_{1}|) \mbox{ vs. } \log (a_{2})$ for  a $k=0$ $\Omega_{o}
=1$ universe. Dashed curve is the solar system constraint and the solid curve
is the nucleosynthesis constraint}
\label{fig1}
\end{figure}

\begin{figure}[h]
\caption{ $\log (|a_{1}|) \mbox{ vs. } \log (a_{2})$ for  a $k=-1$ $\Omega_{o}
=0.1$ universe. Dashed curve is the solar system constraint and the solid 
curve is the nucleosynthesis constraint}
\label{fig2}
\end{figure}

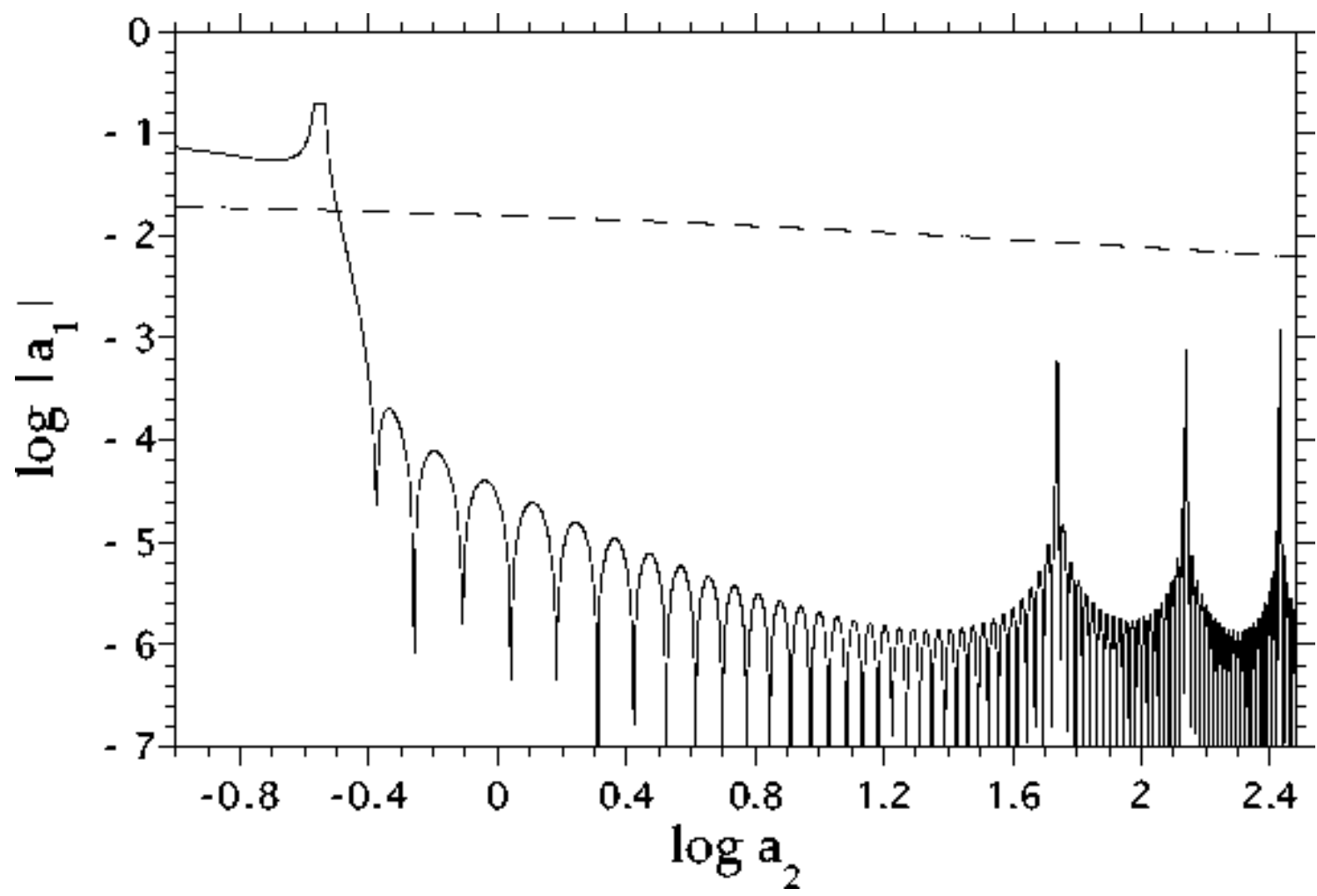
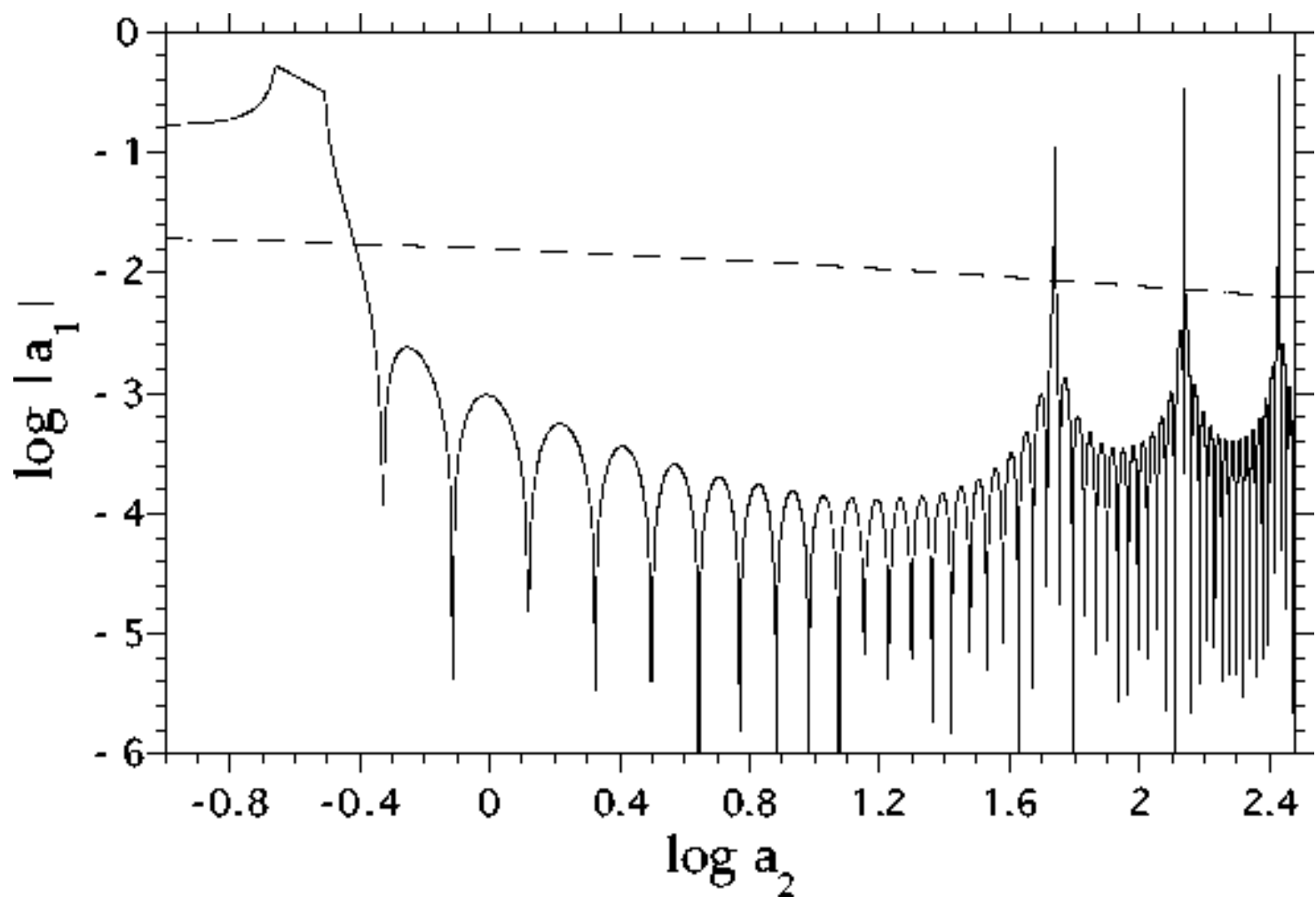
\end{document}